\documentclass[oldversion]{aa} 
\usepackage{txfonts}
\usepackage{graphicx}  
\begin{document}

\title{Response of the warm absorber cloud to a variable nuclear flux in active galactic nuclei}

\author{
L.~Chevallier\inst{1,2}
\and B.~Czerny \inst{2}
\and A.~R\' o\. za\' nska \inst{2}
\and A.~C.~Gon\c{c}alves\inst{1,3}
}
	 
\offprints{\texttt{loic.chevallier@obspm.fr}}

\institute{
Observatoire de Paris-Meudon, LUTH, Meudon, France
\and
Copernicus Astronomical Center, Bartycka 18, 00-716 Warsaw, Poland
\and
Centro de Astronomia e Astrof\'{\i}sica da Universidade de Lisboa, Observat\'orio Astron\'omico de Lisboa, Tapada da Ajuda, 1349-018 Lisboa, Portugal
}

\authorrunning{Chevallier et al.}

\titlerunning{Response of the warm absorber cloud to variable nuclear flux in AGN}

\date{Received 22 May 2006 / Accepted 20 December 2006}

\abstract{Recent modeling of the warm absorber in active galactic nuclei has proved the usefulness of constant total (gas plus radiation) pressure models, which are highly stratified in temperature and density. We explore the consistency of those models when the typical variation of the flux from the central source is taken into account. We perform a variability study of the warm absorber response, based on timescales and our photoionization code \textsc{TITAN}. We show that the ionization and recombination timescales are much shorter than the dynamical timescale. Clouds very close to the central black hole will maintain their equilibrium since the characteristic variability timescales of the nuclear source are longer than cloud timescales. For more distant clouds, the density structure has no time to vary, in response to the variations of the temperature or ionization structure, and such clouds will show the departure from the constant pressure equilibrium. We explore the impact of this departure on the observed properties of the transmitted spectrum and soft X-ray variability: (i) non uniform velocities, of the order of sound speed, appear due to pressure gradients, up to typical values of 100 km s$^{-1}$. These velocities lead to the broadening of lines. This broadening is usually observed and very difficult to explain otherwise. (ii) Energy-dependent fractional variability amplitude in soft X-ray range has a broader hump around $\sim 1 - 2 $ keV, and (iv) the plot of the equivalent hydrogen column density vs. ionization parameter is steeper than for equilibrium clouds. The results have the character of a preliminary study and should be supplemented in the future with full time-dependent radiation transfer and dynamical computations.

}

\maketitle
%

\section{Introduction}

The environment of radio quiet active galactic nuclei (AGN) is still 
poorly understood.
The activity is powered by a black hole but the accretion flow is 
accompanied by an outflow of the material. The inflow is in a form of
an accretion disk which, further out, has a form of a dusty/molecular
torus and the radial velocities of the inflow are not measured directly.
The outflow takes place in a large range of angles around the symmetry axis
but the geometry and the nature of this flow is unclear. The presence 
of the outflow manifests best as Broad Absorption Lines (BAL) of 
optical/UV spectra of some quasars,
Narrow Absorption Lines (NAL) in other quasars or in Seyfert galaxies, 
or as X-ray warm absorbers (WA) in Seyfert galaxies. These absorption
features allow to determine the outflow velocity of the plasma along the
line of sight but any estimates of the spatial distribution of the plasma,
or even of a representative distance of the absorber from
the nucleus, are difficult.

One of the important questions is whether this outflow is roughly a continuous 
wind
with shallow density gradient or a highly clumpy medium with very dense clouds
embedded into low density surrounding plasma. The model of a continuous
wind was proposed by Murray \& Chiang (1995) and subsequently studied e.g.
by Murray \& Chiang (1997, 1998), Proga, Stone \& Kallman (2000, 2004), 
Pietrini \& Torricelli-Ciamponi (2000). On the other hand, the strong X-ray 
irradiation of the outflowing
material may lead to thermal instability (Krolik et al. 1981) thus resulting
in clumpiness of a part of the flow. Therefore, clumpy outflow was advocated
by a number of authors (e.g. Smith \& Raine 1988; Netzer 1993; 
Krolik \& Kriss 1995; Netzer 1996).
A narrow stream of clumpy outflow was 
advocated for by Elvis (2000), while narrow continuous streamers were favored
by Steenbrugge et al. (2005). 
Recently, an advanced model of the clumpy
outflow was developed by Chelouche \& Netzer (2005), based on the previous
extensive studies of the wind acceleration mechanisms.

The issue can be in principle solved directly on the observational grounds.
However,
X-ray data even from grating instruments on board of Chandra or XMM-Newton
telescopes cannot resolve kinematical components narrower than $\sim 100$~km~s$^{-1}$.
High resolution UV data show the presence of absorption lines with the
kinematical widths $\sim 100 $ km s$^{-1}$ (e.g., NGC 3783, Gabel et al. 2003)
or $50 - 60 $ $\sim 100 $ km s$^{-1}$ (e.g., NGC 7469, Scott et al. 2005). These
results are consistent with X-ray data but it is still not clear whether
we probe the same part of the flow with X-ray and UV data. 
Spectral analysis of the X-ray
data suggests that one zone model does not well represent the observations
and at least two photoionization regions are 
required to explain the presence of lines from the gas in 
different ionization states 
(NGC 4051, Collinge et al. 2001;
NGC 5548,  Kaastra et al. 2002;  NGC 3783,  Kaspi et al. 2002, 
Netzer et al. 2003 and
Krongold et al. 2003; Mkn 304, Piconcelli et al. 2004; H0557-385, 
Ashton et al. 2006).  

On the theoretical side, it is not clear whether
strong clumpiness of the medium is possible to achieve and
sustain. Time evolution of the warm absorber medium has not been studied
in great detail so far. Additionally, AGN are known for their 
variability in the X-ray band since
the EXOSAT data. Variable irradiation changes the conditions within the 
cloud and might even destroy the cloud. Also small compact cloud will
respond to variable X-ray flux in a different way than an extended medium.

In the present paper we concentrate on a single warm absorber cloud under
constant pressure. We consider a cloud of considerably large column density
and non-negligible optical depth.
Such a cloud is strongly stratified since the density 
profile is obtained self-consistently from the condition of total pressure 
equilibrium (see Dumont et al. 2002, Chevallier et al. 2006, R\' o\. za\' nska et al. 2006), inducing a few sharp temperature drops due to thermal instabilities.
Magnetic and turbulent pressure effects have been neglected in this study. It is well known that magnetic pressure may remove the thermal instabilities (Bottorff, Korista \& Shlosman 2000), but our cloud, if large enough, will still remain highly stratified with a shallower temperature and density gradient near the location of the previous thermal instability zone. It is the same with turbulent pressure.
Bright irradiated side of the cloud is of low density and highly 
ionized but the density strongly increases inward (by a few orders of 
magnitude) and the medium becomes less ionized as we approach the dark side of 
the cloud.
Such a medium has been used to fit the WA of NGC 3783 (Gon\c{c}alves et al. 2006).
 Here we study the response of the cloud to the varying X-ray flux, with the aim to test the possibility of strong clumpiness in the warm absorber.
 Using the {\it time-independent} code, we estimate the evolution of our cloud at the basis of timescales, determined separately in different parts of our highly stratified cloud. The cloud, initially in total (gaseous + radiative) pressure equilibrium, is disturbed; pressure is no longer constant, and various parts of the cloud expand/contract. We estimate the observational consequences of the cloud departure from equilibrium. Finally we show how the behavior of our cloud, whose distance from the central source is unknown, vary with this distance.
 
In the next section, we recall some generalities about the model and our photoionization code, and we give formulae for timescales which are detailed in the Appendix.
Results are presented in Sect.~3. Section~4 is dedicated to the discussion of the physical and observational implications of these models.

\section{Model}
\label{sect:model}

We study the response of a single warm absorber cloud to a change of the incident X-ray radiation flux, $F$. Using the stationary radiative transfer code, we determine the kinds of equilibrium solutions:
\begin{itemize} 
\item total pressure and thermal equilibrium,
\item only thermal equilibrium.
\end{itemize} 
The first type of solution is the cloud in total pressure equilibrium. It corresponds to a cloud which is in complete equilibrium with the incident flux. The second type of solution is obtained allowing for a change in the irradiating flux but assuming that the timescale of this change, $t_X$, is shorter than the timescale needed to restore the pressure equilibrium. In this case we perform the computations of the new thermal and ionization equilibrium but we retain the original density profile. This second solution will describe the cloud out of pressure equilibrium. The applicability of either the first, or the second solution depends on the cloud location and the variability properties of the nucleus. 

The radiative transfer used in both cases is the code TITAN (Dumont et al. 2000, 2003). This 1-D code works in plane-parallel symmetry but uses the Accelerated Lambda Iteration (ALI) method to provide fully-consistent multi-angle radiative transfer computations for both continuum and lines (Rybicki \& Hummer 1991, and references therein). Using an accurate radiative transfer method is important in such a problem, where radiation pressure variation within the cloud plays an important role, especially in the determination of the location of temperature drops due to thermal instabilities. 
We consider about 1000 spectral lines, mainly in the X-ray range of the 10 most abundance species in the universe, i.e. H, He, C, N, O, Ne, Mg, Si, S and Fe. Throughout the paper we use cosmic abundances (Allen 1973), which are constant in the cloud.

The complete equilibrium state is characterized by the local density at the irradiated face of the cloud, $n_o$, the hydrogen column density of the cloud, $N_H$, and the properties of the incident radiation (a power-law of index 1 from 10 eV to 100 keV, and the ionization parameter  $\xi = L/n_o R^2$, where $L$ is the bolometric luminosity of the source, $R$ the distance of the medium from the source). The density profile is determined consistently from the condition of the total pressure equilibrium. The off-equilibrium state is characterized by a new value of the ionization parameter $\xi$ but the density profile is taken from the previous equilibrium solution.

We first analyze the timescales for various processes within the cloud in full equilibrium, in order to establish self-consistency of the cloud response. 

The light travel time is given by
\begin{equation}
\label{eq:tlight}
t_{light} = {H \over c},
\end{equation}
where $H$ is a geometrical size of a whole cloud, or of a zone under consideration. In our case this timescale is relevant since the cloud is optically thin to the continuum. This low (but not negligible) optical depth is caused by significant ionization of the medium even at the dark side of the cloud.

If the irradiation changes, the whole volume of the cloud responds to the decrease/increase in photon flux. Therefore, generally we do not have here well defined ionization fronts, as in the case of an expanding HII region around a star (e.g. Osterbrock 1974). The relevant description is provided, instead, by the ionization and recombination timescales. 
The ionization $t_{ion}$ and recombination $t_{rec}$ timescale for each ion is a local quantity within a cloud. However, we can introduce timescales for specific zones, or a whole cloud, by analyzing which ions dominate in the various zones and which transitions are close to 
equilibrium (see the Appendix). We obtain the following estimates
\begin{equation}
\label{eq:tion}
t_{ion} = 6 \times 10^{-3} F_{16}^{-1}~[\mathrm{s}],
\end{equation} 
and
\begin{equation}
\label{eq:trec}
t_{rec} = 0.2 n_{12}^{-1}~[\mathrm{s}],
\end{equation}
where $F_{16}$ is the incident bolometric flux measured in units of $10^{16}$~erg s$^{-1}$ cm$^{-2}$, and $n_{12}$ is the cloud hydrogen density at the surface in units of $10^{12}$~cm$^{-3}$.

The local thermal timescale is defined as
\begin{equation}
\label{eq:t_th}
t_{th} = {kT \over n_e \Lambda} = 12 T_6 n_{12}^{-1} \Lambda_{23}^{-1}~[\mathrm{s}],
\end{equation}
where $k$ is the Boltzmann constant, $\Lambda_{23}$ is the cooling function in units of $10^{-23}$~erg~cm$^3$~s$^{-1}$, $T_6$ is the temperature in units of $10^6$~K, and the 
electron density $n_e \sim 1.18 n_H$, where $n_H$ is hydrogen density.
Isothermal sound speed velocity is locally calculated as
\begin{equation}
c_s = \sqrt{2 k T \over m_H} = 130 T_6^{1/2}~[\mathrm{km~s}^{-1}],
\label{eq:sound_speed}
\end{equation}
where $m_H$ is the hydrogen atomic mass and the factor 2 accounts for mean molecular weight in an ionized medium.
Eq.~\ref{eq:sound_speed} is relevant when taking into account only the gaseous pressure for a perfect gas.
With such a stratified cloud, the dynamical timescale (i.e. time for restoration of pressure equilibrium) of a cloud zone of a geometrical size $\Delta z = z_2 - z_1$ is
obtained through integration over $z$
\begin{equation}
t_{dyn} = \int_{z_1}^{z_2}{dz \over c_s(z)},
\label{eq:dyn}
\end{equation}

We generalize the results for a single cloud to a whole family of clouds at various distances from the nucleus.
The initial configuration (i.e. the variation of all quantities in optical depth scale) practically does not depend on $n_o$ (with values between $10^5$ to $10^{12}$ cm$^{-3}$; see R\' o\. za\' nska et al. 2006) for a given $\xi$ and $N_H$. Since a change of the local density at the illuminated surface of the cloud is equivalent to a change of distance $R$ from the nucleus we can take our cloud as representative for a whole family of self-similar clouds at various distances. So our model can be applied to several distances $R$, with appropriate dependence of $n_o$ on the radius $R$.
We have $F(R) \propto R^{-2}$, $T$ as the function of the hydrogen column remains the same, independently from the distance, and $n$ scales with $n_o$. We request $n_o(R) \propto R^{-2}$, thus $H (R) \propto R^2$. Consequently all timescales defined in this section are proportional to $R^2$.

\section{Results}

\subsection{Initial state of the cloud}
\label{sect:initial}

We model first the cloud in complete total pressure equilibrium. For this reference cloud, we choose $\xi = 10^4$~erg~cm~s$^{-1}$, $n_o=10^7$ cm$^{-3}$. 
We choose the largest $N_H\sim3\times 10^{23}$~cm$^{-2}$ allowed for the constant pressure cloud for such an incident spectrum (Chevallier et al. 2006). 
This choice of the column-density is consistent with computations of the cloud sizes by Torricelli-Ciamponi \& Courvoisier (1998), which take into account the thermal conduction. If we apply their expression for the cloud core scaling (Eq.~5), taking their cloud dimensionless size $x=3\times 10^7$ from their Figs.~4 and 5, and take the core temperature $3\times 10^4$~K (i.e. the smallest temperature in our reference cloud), we obtain the expression for the column-density of the cloud $N_H\sim2\times 10^{23}$~cm$^{-2}$, independent from the core number density.  The internal kinematics is set as random Gaussian movements with dispersion 100 km s$^{-1}$, whose only effect is to increase the line width by Doppler effect.

\begin{figure}
\resizebox{\hsize}{!}{\includegraphics{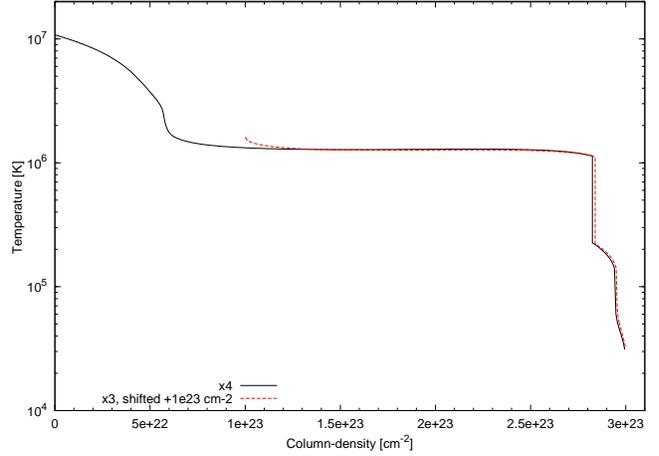}}
\caption{The temperature profile of the cloud at the initial state $\xi=10^4$~erg~cm~s$^{-1}$ (red). We also represent the case more typical of WA: $\log\xi=3$, for which maximal column-density is $2\times 10^{23}$~cm$^{-2}$ (magenta). This curve is shifted in column-density by a value $10^{23}$ cm$^{-2}$ in order to compare with the $\log\xi=4$ case. Curves are not distinguishable.}
\label{fig:temp_special}
\end{figure}

The temperature profile across the cloud as a function of the hydrogen column-density measured from the cloud surface is shown in Fig.~\ref{fig:temp_special}. For such abundances, electron density is $n_e\sim 1.18\ n_\mathrm{H}$ and total density is $n\sim 2.265\ n_\mathrm{H}$). The cloud develops four characteristic zones: 
\begin{itemize}
\item low density high temperature extended zone (hot), 
\item intermediate zone (intermediate),
\item high density low temperature outer zone (outer),
\item thin zone of highest density and lowest temperature (end). 
\end{itemize}

The last zone is geometrically very thin, as can be seen in Table~\ref{tab:zones} but the hydrogen column density of this zone is significant because of the high density. For the considered cloud, the ratio of densities at the cold and hot surfaces is $\sim 10^3$, whatever the value of density $n_o$ between $10^5$ and $10^{12}$ cm$^{-3}$. The geometrical size of the reference cloud is $5 \times 10^{15}$ cm, smaller by a factor of 6 than for a constant density cloud with the same column density and the same density at the cloud surface. The light travel time across the cloud is $1.7 \times 10^5$ s.
The typical value of $\xi$ of warm absorbers, estimated from the observational data, is rather $\sim 1000$ or lower, and such values are met in the interior of our reference cloud (i.e. intermediate zone, see Table~\ref{tab:zones}) where the density is roughly ten times the density at the surface of our medium.
Figure.~\ref{fig:temp_special} confirms that models with $\log\xi=3$ and 4 behave similarly. For our $\log\xi=3$ model, $N_H \sim 2\times 10^{23}$~cm$^{-2}$, which is now the maximum column density. In order to compare these temperature profiles, we have shifted the solution for $\log\xi=3$ by $10^{23}$ cm$^{-2}$ along the column-density axis. Temperature profiles are almost the same, apart that the intermediate zone for the $\log\xi=3$ case is 25\% less than for the $\log\xi=4$ case.

As ionization or recombination timescales are specific of each ion, we analyzed the ionization structure of a whole cloud (see the Appendix and Table~\ref{tab:maxion_zones}).
The longest ionization timescale is equal to $8 \times 10^3$~s (for ion Fe{\sc xxvi} in the hot zone). Processes representative for the intermediate, outer and end zones are connected with S{\sc xvi}, O{\sc viii} and C{\sc vi}, respectively.
The representative recombination timescale for a whole cloud is $2 \times 10^4$ s (again for Fe{\sc xxvi} in the hot zone). Representative ions for the recombination timescale in the intermediate, outer and end zones are Si{\sc xiv}, O{\sc ix} and C{\sc vii}, respectively.

\begin{figure}
\resizebox{\hsize}{!}{\includegraphics{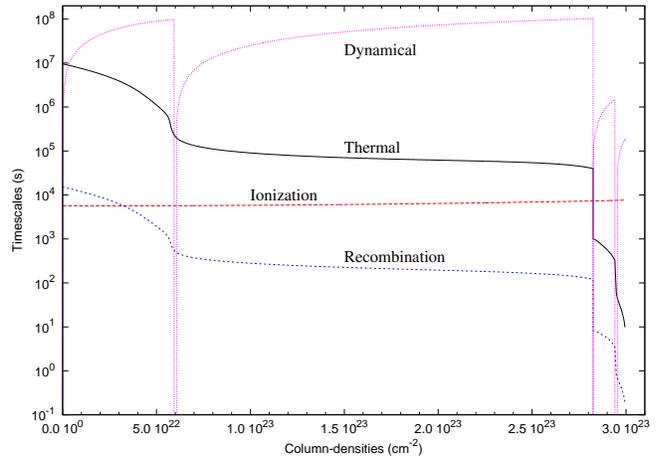}}
\caption{The dependence of the various local timescales (s) across the cloud (as a function of the column-density): ionization timescale, recombination timescale, thermal equilibrium timescale, hydrostatic equilibrium timescale (cumulative through the slab, classical perfect gas). As ionization or recombination timescales are specific of each ion, we choose to represent Fe{\sc xxvi}, which represent the longest timescale (only in the hot zone for recombination, leaders are Si{\sc xv}, O{\sc ix} and C{\sc vii} for the intermediate, outer and end zones, respectively; see Tables~\ref{tab:zones} and \ref{tab:maxion_zones}).}
\label{fig:time}
\end{figure}

The profiles of the timescales calculated locally across the cloud are shown in Fig.~\ref{fig:time}. 
We see that photoionization timescale is roughly constant through the slab but the recombination timescale is monotonically decreasing through the cloud since it is inversely proportional to the electron density $n_e$ according to the scaling formula (see Eq.~\ref{eq:trec}).
The actual recombination timescale calculated numerically across the cloud (see the Appendix) depends also on the temperature (not linearly) and decreases even faster. The recombination timescale is more shallow if we consider various ions as representative for different zones (see Table~\ref{tab:zones}). 

The overall thermal timescale depends both on the temperature and the density, so it decreases with the distance from the cloud surface even more rapidly than the recombination timescale (see Table~\ref{tab:zones}), and roughly follows the numerically obtained recombination timescale for a single ion. Its maximum value of $10^7$, at the hot surface, is very high, much longer than ionization or recombination timescales. We have checked that $\Lambda_{23} \sim 1$ whatever the depth for our model, including the surface layers. In other zones the thermal timescale is shorter, so most ions reach equilibrium in the timescale appropriate for the intermediate zone ($2\times 10^5$~s), and we adopt this value as the longest characteristic thermal timescale of the cloud. However, we must stress that in the case of the hot zone, where highly ionized iron spectral features form, the full thermal equilibrium is hard to achieve. The temperature rises/falls slowly since both the atomic heating/cooling and the Compton heating/cooling are not very efficient.
The dynamical timescale of the whole cloud obtained through integration is equal $2 \times 10^8$ s. Both in the hot and in the intermediate zone these timescales, calculated separately, are long (see Table~\ref{tab:maxion_zones}). In Fig.~\ref{fig:time} we plot the cumulative timescales for each of the zones separately, starting from the beginning of a zone.

\begin{table}
\centering
\begin{tabular}{@{}ccccc@{}}
\hline
&  hot & intermediate & outer & end \\
\hline
$\Delta z$ [cm]  & $3.4\ 10^{15}$ & $1.6\ 10^{15}$ & $8.4\ 10^{12}$  & $9.9\ 10^{11}$ \\
col. [cm$^{-2}$] & $5.8\ 10^{22}$  & $2.3\ 10^{23}$  & $1.3\ 10^{22}$ & $5.3\ 10^{21}$ \\
$T$  av. [K] &  $8.0\ 10^6$  & $1.3\ 10^6$  & $1.9 \ 10^5$  & $4.8 \ 10^4$ \\ 
$n_\mathrm{H}$ av. [cm$^{-3}$] & $1.7\ 10^7$ & $1.4\ 10^8$ & $1.5\ 10^9$ & $7.2\ 10^9$ \\
$c_s$ [km~s$^{-1}$] & $370$& $150$ & $57$ & $28$ \\
$t_{light}$ [s] & $1.2\ 10^5$ & $5.4\ 10^4$ & 290 & 34 \\
$t_{ion}$ [s] & $6\ 10^3$ & $10^3$ & 100 & 30 \\
$t_{rec}$ [s] & $2\ 10^4$ & $2\ 10^3$ & 100 & 20 \\
$t_{th}$ [s] & $10^7$ & $2\ 10^5$ & $10^3$ & 60 \\
$t_{dyn}$ [s] & $9.3\ 10^7$ & $1.1\ 10^8$ & $1.5\ 10^6$ & $3.5\ 10^5$ \\
\hline
\end{tabular}
\caption{Characteristics of the four zones in this medium, and all timescales considered in this study (see the text). $t_{light}$ and $t_{dyn} = \Delta z/c_s$ are computed zone by zone ($t_{dyn}=2\times 10^8$~s for the whole cloud).}
\label{tab:zones}
\end{table}

\subsection{Cloud states and conditions for the self-consistency of the cloud response description}

The X-ray flux of AGN nuclei is strongly variable.
Since the timescales corresponding to ionization/recombination equilibrium and 
to the dynamical equilibrium differ by a few orders of magnitude, a cloud is not always likely to achieve a complete equilibrium. We will now analyze the conditions of complete, or partial equilibrium.
 
For this purpose, we must compare the cloud internal timescales with the timescales of the intrinsic variability of the X-ray incident flux. AGN do not have any specific timescales but the power spectrum in Seyfert galaxies is well represented by a power law with the two breaks (e.g. Markowitz et al. 2003). Most of the power is between the high and the low frequency breaks, which depend on the central black hole mass. We can consider these frequency breaks of the power spectrum in an AGN as a characteristic range of timescales.  We adopt $t_X^{short}$ and $t_X^{long}$ as the shortest and longest timescales of variation, respectively. With a black hole mass of $10^7 M_{\odot}$, the appropriate timescales for regular Seyfert 1 galaxies are $t_X^{short} = 6 \times 10^5$ s, and $t_X^{long} = 6 \times 10^7$ s (Markowitz et al. 2003, Papadakis 2004), and the short timescale is by a factor 20 shorter for Narrow Line Seyfert 1 galaxies. For illustrative purposes, we further adopt $t_X^{short} = 10^5$~s and $t_X^{long} = 10^7$~s.

The full thermal and pressure equilibrium within the cloud is achieved if the light crossing time, the longest of the ionization, recombination or thermal timescales, the dynamical time of the cloud are much shorter than  the shortest X-ray variability timescale, i.e. $\max( t_{light}, t_{th}, t_{ion}, t_{recomb},t_{dyn}  ) \ll t_X$. On the other hand, the thermal equilibrium solution with an unchanged density profiles requires  $\max( t_{light}, t_{th}, t_{ion}, t_{recomb}) \ll t_X \ll t_{dyn}$.

In our reference cloud the second set of conditions is roughly fulfilled. The ionization and recombination timescales are short for all important transitions. The thermal timescale in the intermediate zone is comparable to $t_X^{short}$, only the thermal timescale in the hot zone is significantly longer which may effect the hydrogen-like iron absorption feature. The shortest dynamical timescale in the last zone is still longer than $t_X^{short}$, so the model is appropriate for short timescale variability. Longer systematic trends, like those in  $t_X^{long} = 10^7$ s, should be followed by computing new pressure equilibrium at the back side of the cloud, in the outer and end zones but retaining the old density profile in the hot and intermediate zone. We postpone the computations of such hybrid models for the future. 

Summarizing, our new equilibrium is fully self-consistent for an increase in the X-ray flux lasting up to $t_X = 3 \times 10^5$ s (i.e. the shortest dynamical timescale, see Table~\ref{tab:zones}) if the density profile is left unchanged. Results obtained for longer timescales can only serve as indicators of the trends.

\subsection{Response of the cloud structure to the change of irradiation flux}
\label{sect:response}

\begin{figure}
\resizebox{\hsize}{!}{\includegraphics{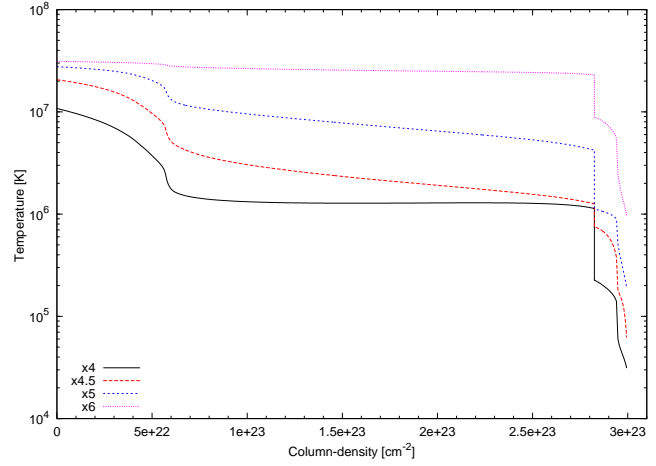}}
\caption{The temperature profile of the cloud at the initial state $\xi=10^4$~erg~cm~s$^{-1}$ and
after an increase of the irradiation to $\xi=3 \times 10^4$, $10^5$ and $10^6$~erg~cm~s$^{-1}$, but before a new pressure equilibrium is reached. For all those cases, we have neglected the turbulent pressure.}
\label{fig:temp1}
\end{figure}

We therefore compute a set of new cloud equilibrium solutions by increasing the incident flux and preserving the density profile from the pressure equilibrium state described in Sect.~\ref{sect:initial}. 
Figure~\ref{fig:temp1} shows the temperature profiles of the cloud for which the increase of flux is by a factor 3, 10 and 100. 
Naturally, an increase in the X-ray flux results in a higher surface temperature. The depth of the first, hottest zone does not change but the temperature of the intermediate zone rises. An increase in X-ray flux by a factor 100 removes the intermediate zone completely. The two last colder zones always exist although for the highest irradiation flux the temperature of the back side of the cloud rises up to $10^6$ K. However, the rise of the incident flux by a factor of 3 already brings up the temperature of the outer zone almost close to the temperature of the intermediate zone.

\subsection{Response of the cloud velocity field to the change of irradiation flux}\label{sec:velocity}

\begin{figure}[t]
\resizebox{\hsize}{!}{\includegraphics{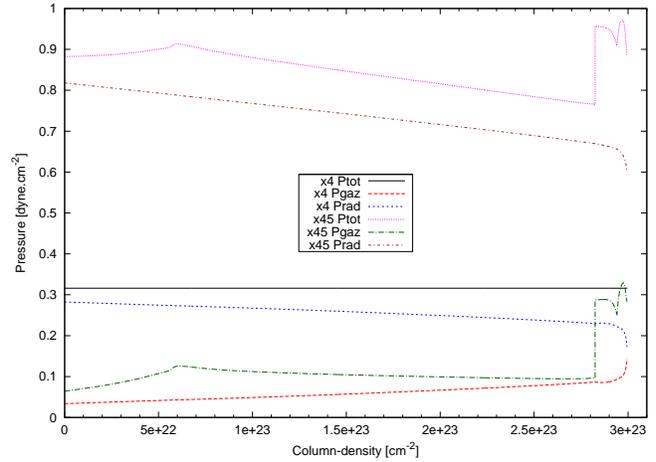}}
\caption{Variation with the column-density of gaseous, radiation and total pressure for the $\log\xi=4$ constant total pressure case, and under  illumination $\log\xi=4.5$ for a fixed density profile.}
\label{fig:pressure_variation}
\end{figure}

After the change in the incident flux, the cloud is not in the pressure equilibrium any more. Pressure gradients appear which lead to local acceleration of the gas within the cloud and subsequent expansion. Figure.~\ref{fig:pressure_variation} shows the gaseous, radiative and total pressure for the initial state $\log \xi = 4$ and the final state $\log \xi = 4.5$. The most prominent feature is the huge positive pressure gradient at the transition between the intermediate and outer zones. Therefore, a shock may form at this location as the material strain to recover the dynamical equilibrium. It is not a classical ionization front since the whole cloud is in a new ionization equilibrium after a relatively short time $t_{ion}$. This is a different type of shock which forms due to the presence of the density gradient in the original model, previously supported by the appropriate temperature gradient. This support is removed by the sudden rise of the temperature on the cold side. Since the discontinuity in the total pressure is not very large (a factor of 1.25 in case of the flux rise by a factor 3, see Fig.~\ref{fig:pressure_variation}) the expected Mach number, $\cal M$, of an isothermal shock is low (${\cal M}^2 = 1.25$), a relatively weak shock forms and propagates with a speed close to the speed of sound ${\cal M} c_s \sim 120$~km~s$^{-1}$ in this zone, as we only have to consider the highest temperature $T\sim 7\ 10^5$~K ahead the thermal instability zone (see Fig.~\ref{fig:temp1}). Subsonic motions due to the lack of pressure balance will also appear in other parts of the cloud.

Due to the very different values of the velocity in our cloud, we thus expect that lines forming in these transition zones may have a very complex profile: part of the line may be narrow, and part of the line profile may be broad. The presence of the significant velocity gradient means that the assumption of a constant turbulence across the cloud may not be justified but the computations of the radiative transfer with variable velocity field is beyond the scope of the present paper.

\subsection{Clouds at various distances from the nucleus}\label{sec:distance}

The choice of $n_o = 10^7$ cm$^{-3}$ and $\xi=10^4$ corresponds to the choice of the distance for a specific choice of the bolometric luminosity $L$ of the nuclear source.
The domain of observed luminosities range form $10^{41}$~erg~s$^{-1}$ (e.g., NGC 4051) up to $10^{47}$~erg~s$^{-1}$ (e.g., quasars).
We adopt $L=10^{45}$ erg s$^{-1}$ cm$^{-2}$ in the further calculations: the cloud is located at the distance of $10^{17}$~cm from the nucleus. All those results scale as $L/R^2$, and any value of $L$ can be chosen.

The results obtained for a single cloud can be generalized to a family of clouds located at various distances from the nucleus adopting scaling laws based on the very weak dependence of the results for the radiative transfer on the adopted value of the number density at the cloud surface, $n_o$ (see Sect.~\ref{sect:model}).

We have checked the underlying assumption by calculating a set of clouds for $\xi = 10^4$~erg~cm~s$^{-1}$, $N_H = 3 \times 10^{23}$ cm$^{-2}$ and with values of $n_o$ in the range from $10^5$ to $10^{11}$ cm$^{-3}$. The temperature profiles measured as functions of the optical depth were almost identical. The density profiles as functions of the optical depth had very similar shape and differed just in the normalization. The ratio of the density at the outer edge of the cloud to $n_o$ was almost constant, only slowly decreasing from $1.5\times 10^3$ to $10^3$ with the increasing density.

Therefore, at the basis of the numerical results for a single cloud we consider now a family of clouds for $\xi = 10^4$~erg~cm~s$^{-1}$, $N_H = 3 \times 10^{23}$ cm$^{-2}$, and the bolometric luminosity of the nucleus $L = 10^{45}$ erg s$^{-1}$ cm$^{-2}$. The condition of constant $\xi$ in this case implies that the number density at the cloud illuminated surface is $n_o(R) / R_{17}^{-2} = 10^7$~cm$^{-3}$, where $R_{17}$ is the distance in units of $10^{17}$ cm, and it is by a factor of $10^3$ higher at the dark surface. Other quantities defined in Sect.~\ref{sect:model} become $H(R) /  R_{17}^{2} = 5 \times 10^{15}$~cm, $t_{light}(R) / R_{17}^2 = 1.7 \times 10^5$~s, $t_{ion}(R) / R_{17}^2 = 8.0 \times 10^3 - 30$~cm, $t_{rec}(R)  / R_{17}^2 = 2 \times 10^4 - 20$~s, $t_{th}(R) / R_{17}^{2} = 2 \times 10^5 - 60$~s, and $t_{dyn}(R) / R_{17}^{2} = 10^8 - 4 \times 10^5$~s.

\begin{figure}
\resizebox{\hsize}{!}{\includegraphics{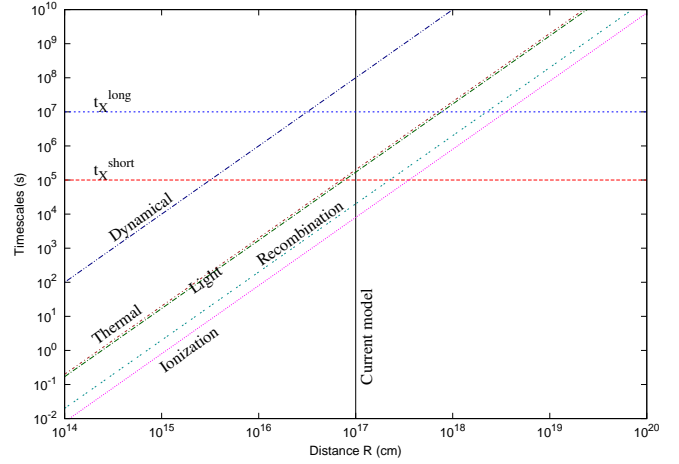}}
\caption{The dependence of the maximum value within the cloud of various timescales as a function of the distance from the nucleus for a source with luminosity $L=10^{45}$~erg~s$^{-1}$. We represent the bounds of the characteristic range of variability of the source, i.e. the the shortest timescale $t_X^{short}$ and longest timescale $t_X^{long}$, light travel time, ionization timescale, recombination timescale, thermal timescale and dynamical timescale. This value of the thermal timescale excludes the hot zone, where it is up to 50 times longer (see Table~\ref{tab:zones}). Vertical line indicates the current model presented here (equivalent to $R=10^{17}$ cm).}
\label{fig:skale}
\end{figure}

Using these expressions and a variability timescale of the source between $10^5$~s and $10^7$~s, we can estimate the kind of equilibrium reached by our medium as a function of the distance. The results are shown in Fig.~\ref{fig:skale}, where several regions, as a function of the distance of our cloud from the nucleus, can be defined. Our reference cloud (i.e. hydrogen surface density $n_o=10^7$~cm$^{-3}$) is marked with the vertical line. Horizontal lines denote the bounds of the characteristic range of variability timescales of the source, for a given central luminosity, $L=10^{45}$~erg~s$^{-1}$. Note that those results depend only on the value of the ionization parameter, $\xi$, and the cloud column density, $N_H$, and this figure can be scaled with various values of the luminosity $L$.  
We see from Fig.~\ref{fig:skale} that in general we can have four basic types of clouds:
\begin{itemize}
\item full (dynamical + thermal) equilibrium, $R < 3 \times 10^{16}$ cm
\item thermal equilibrium, $ 3 \times 10^{16}$ cm $ < R < 3 \times 10^{18}$ cm
\item no ionization equilibrium, $ 3 \times 10^{17}$ cm $ < R < 5 \times 10^{18}$ cm 
\item clouds at $R > 5 \times 10^{18}$ cm
\end{itemize}
The innermost clouds should be described using full equilibrium computations. The second group of clouds shows the departure from the dynamical equilibrium, but their ionization equilibrium is preserved. Their ionization state should be roughly correlated with the X-ray flux measured at a give moment, and they can be represented as described in Sect.~\ref{sect:response}. In particular, a shock forms and propagates at velocity $\sim 120$~km~s$^{-1}$ whatever the distance $R$, as the variability timescale $t_X$ and the extent of the outer zone are both higher than the time and displacement, respectively, of the matter when the sound speed is reached (see Sect.~\ref{sec:velocity}). Clouds further up are so extended that separate parts of the cloud responds to different continuum, as their light travel time is comparable to the timescales of the X-ray source variability. Only complete time-dependent computations of the radiative transfer, with the adopted light-curve, are appropriate to analyze such clouds, and at present we are unable to perform such computations. For the last group of very distant clouds, the cloud size is so extended that the assumption of a single cloud breaks down and we deal with an extended continuous wind.  

\section{Discussion}

In this paper we have considered the complexity of the response of the warm absorber to the variable X-ray incident flux. In order to assess the cloud response we mostly concentrated on comparison of two states of a single warm absorber cloud: an initial irradiated cloud at pressure equilibrium, and a cloud {\it suddenly} irradiated by a three times higher flux i.e. that cloud could reach new thermal and ionization equilibrium under enhanced irradiation but the time was too short to reach new pressure equilibrium. The model can be used against the presently available and future observational data.

Below we give a few exemplary applications of our representative cloud results, not yet focused at fitting a specific object. More precise modeling should use a family of solutions. 

\subsection{Observational applications}

\subsubsection{Soft X-ray variability}

Enhanced irradiation leads to higher ionization of the cloud and lower opacity, particularly in the soft X-ray band. Indeed, observations based on RXTE monitoring show an enhancement of the energy-dependent fractional variability amplitude below 5 keV (Markowitz et al. 2003). Best studied case of a Seyfert 1 galaxy MCG-6-30-15 shows enhanced variability in 0.2--4 keV band approximately equal to 30\% (Fabian et al. 2002, Ponti et al. 2004). The variability below 0.2 keV and above 5 keV must be intrinsic, as it reflect the changes in the level (and possibly slope) of the continuum but the excess 0.2--4 keV may be due to the opacity changes in the warm absorber (e.g. Sako et al. 2002). Such an effect on this object has been studied using constant density clouds (Gierlinski \& Done 2006).

\begin{figure}
\resizebox{\hsize}{!}{\includegraphics{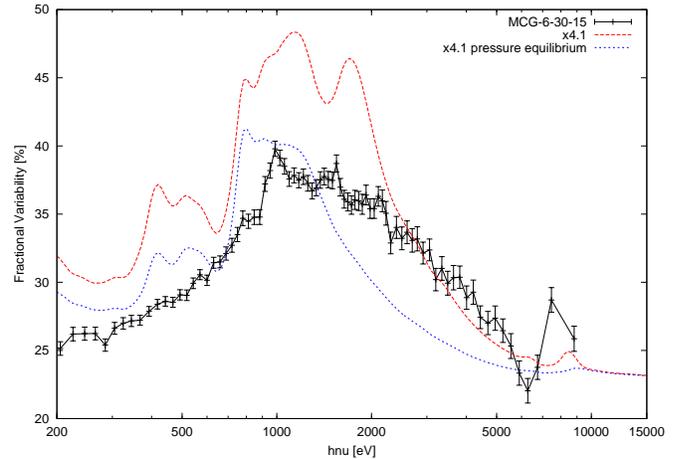}}
\caption{The fractional variability amplitude (here in \%, defined as $2 |F_1-F_2| / (F_1+F_2)$) calculated using the $\log\xi=4$ spectrum as a reference, for two models of increasing flux to $\log\xi=4.1$, the first one as described in this study and the second one in total pressure equilibrium. Those quantities are shown for a wide energy range: 200 eV to 15 keV. Some difference is seen in the width of the horizontal zone between 1 and 2 keV, and for the features around 7 keV (iron $K\alpha$ line region). Spectral resolution is 10 (Gaussian shape). The overall level of the amplitude reflect the assumption about the increase in the flux but the energy-dependent features can be compared to the features in the observed fractional variability amplitudes. The exemplary data points represent a 100 ksec XMM observation of MCG-6-30-15 (Goosmann et al. 2006). 
}
\label{fig:spec_ratio}
\end{figure}

Using our density-stratified clouds, we have calculated the ratio of the two spectra (models with $\log\xi=4$ and $\log\xi=4.5$). Since the change of the flux by a factor 3 is rather extreme as compared to the observed variability in MCG-6-30-15, we only consider the model with $\log\xi=4.1$ (i.e. increase of $\sim 26$\% of the incident flux) without adjustment of the pressure and another one with $\log\xi=4.1$ in constant pressure equilibrium.
In Fig.~\ref{fig:spec_ratio} we show the ratios of the two spectra plotted with the resolution usually used from the computation of the fractional variability amplitude. We also show the fractional variability amplitude calculated from one of the data sets obtained for MCG-6-30-15 (XMM observation in 2000, rev. 303;  see Goosmann et al. 2006). Indeed, there is a strong enhancement of the variability at 1--2 keV. The maximum value of 50\% of the fractional variability for the $\log\xi=4.1$ case is close from the MCG-6-30-15 case. This suggests that the variable ionization of the warm absorber is a promising contributor to the overall variability. If so, such enhanced variability in the soft X-ray band peaking at 1 keV should be seen only in the sources which show the presence of the warm absorber. Our model shows slightly too much absorption for this data set, and a better fit would be reached with a lower column-density, for the same value of $\xi$. It is also interesting to point out that the fractional variability enhancement obtained without the pressure equilibrium extends more into high energies, and seems to fit better the data. This would suggest the warm absorber distance larger than $3 \times 10^{15} $ cm since the medium had no time to reach the dynamical equilibrium (see Sect.~\ref{sec:distance}).

\subsubsection{Observational determination of the column density}

\begin{figure}
\resizebox{\hsize}{!}{\includegraphics{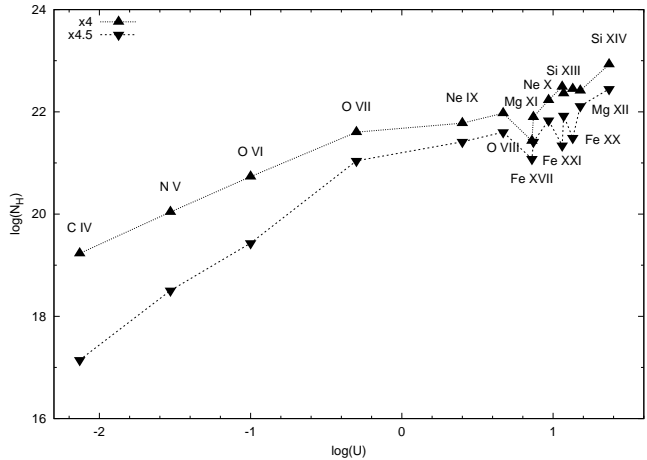}}
\caption{Equivalent hydrogen column-density vs. ionization parameter for some ions selected in our reference cloud in its initial state ($\log\xi = 4$) and width the increased illumination ($\log\xi = 4.5$). Apparent systematic increase reproduces the trend seen in the data of Scott et al. (2005).}
\label{fig:ion_relation}
\end{figure}

An interesting trend was found of increasing equivalent column density of the absorber with an increase of the ionization parameter for the species (NGC 5548, Steenbrugge et al. 2005; NGC 4051, Ogle et al. 2004; NGC 7469, Scott et al. 2005). It is therefore interesting to see whether such a trend is explained by our constant pressure clouds. Therefore, we selected the ions studied by Scott et al. (2005) and performed a similar analysis, i.e. we used the ion column densities from our simulations (for the initial cloud and the cloud with increase irradiation, out of pressure equilibrium) and inverted them to equivalent hydrogen column using only abundance values. The results are shown in Fig.~\ref{fig:ion_relation}. We see that the trend seen in the data is in a natural way explained by our model. The ratio of almost five orders of magnitude between the results for Fe{\sc xx}, Mg{\sc xii}, Si{\sc xii}, Si{\sc xiv} and C{\sc iv} is well explained by the model out of dynamical equilibrium. On the other hand, our column densities are lower than these obtained by Scott et al. (2005). Extensive modeling with the use of other cloud parameters will be necessary to see whether the specific values characteristic for NGC 7469 can be accommodated within our model. 

\subsubsection{Line widths}

The predicted line widths depend on the cloud location, for a fixed nuclear source 
properties. Cloud located closer in reach equilibrium much faster and the velocity
gradient is smaller. Clouds located much further away will also show smaller velocity
gradients since they are too extended to react to the variable radiation flux. 
Therefore, the measurement of the line width will provide an independent limit to
the cloud distance, and an additional test of the constant pressure cloud models.

We illustrate this issue in Table~\ref{table:widths}. We take the measured kinematical
width of absorption lines measured in the X-ray spectrum of NGC 7469 by Scott et al. (2005) and we compare them with our estimates based on the location of the ions.

Measured width in the X-ray band are always larger and given with significant 
errors. It is probable that many lines are actually blends and the
resolution in X-ray spectra is not yet satisfactory to test the model. Our model predicts many more strong absorption lines in this energy band. Lines are measured 
much more precisely in the UV part of the spectrum so in the future out study will be
extended into UV spectral region. The available measurements for NGC 7469 UV data
cannot be used immediately for our purpose since line intensities and shifts were measured for the
lines widths fixed at the same value for all narrow absorption components (61 km/s and 
45 km/s for two kinematical components; Scott et al. 2005).

\begin{table}
\begin{tabular}{lllrr}
\hline
Wavelength & absorption  & zone         &     model  & measured \\
  \\A     &    line              &              &       [km/s]     &   [km/s]   \\
\hline
 6.18..   &    Si{\sc xiv}    &     intermed   &           150     &  $620 \pm 690$ \\
 6.65...  &    Si{\sc xiii}   &     intermed   &           150     &  $1350 \pm 900$ \\
 8.42...  &    Mg{\sc xii}    &     intermed   &          150     &  $500 \pm 160$ \\
 9.17...  &    Mg{\sc xi}     &     intermed/outer  &     150     &  $680 \pm 410$ \\
 11.55... &    Ne{\sc ix}     &     outer             &     60     &  $1020 \pm 520$ \\
 12.13... &    Ne{\sc x}      &     intermed/outer  &     150     &  $430 \pm 270$ \\
 12.28... &    Fe{\sc xxi}    &     intermed      &        150     &  $890 \pm 570$ \\
 12.82... &    Fe{\sc xx}     &     intermed         &       150     &    $1070 \pm 340$ \\
 13.45..  &    Ne{\sc ix}     &     outer            &        60     &  $1050 \pm 490$ \\
 15.01... &    Fe{\sc xvii}   &     outer             &       60     &  $650 \pm 320$ \\
\hline
\end{tabular}
\caption{Kinematical widths of the X-ray lines estimated from our cloud model in comparison with the line widths measured by Scott et al. (2005). Third column give the dominant location of the ion (lines identified by Scott et al. 2005 all come either from the intermediate or from the outer zone).}
\label{table:widths}
\end{table}

\subsection{Further improvements in modeling}

Our approach to modeling the time-dependent response of the cloud to irradiation was a
first step in this direction. It is well known that the problem is complicated, and the best
studied classical example is the time-dependent evolution of the expanding $H^+$ region around a newly born star (see Osterbrock 1974). Unfortunately, our situation is more complex since we do not deal with a single ionization front, and simple analytical estimates do not apply so easily to our case. However, further improvement of the description of the cloud dynamics is necessary.

Particularly important and difficult problem appears if clouds are at large distances from the center. Blustin et al. (2005) argued that the warm absorber is located in most sources at distance $\sim 1$~pc. Our cloud at such a distance is far from radiative equilibrium. In this case time-dependent radiation transfer computations seem to be necessary.

\begin{acknowledgements}
We thank to S. Collin, C. Done, T. Kallman, F. Nicastro and I. Shlosman for helpful discussions.
We also thank A.-M. Dumont for providing us some help with the code {\sc TITAN}. 
A.~C. Gon\c{c}alves acknowledges support from the {\it {Funda\c{c}\~ao para a Ci\^encia e a Tecnologia}}, Portugal, under grant no. BPD/11641/2002.
Part of this work was supported by grant 1P03D00829 of the Polish State Committee for Scientific Research, and the Laboratoire Europ\'een Associ\' e Astrophysique Pologne-France.
\end{acknowledgements}

\section{Appendix}

In order to obtain the ionization timescale $t_{ion}$ of a whole 
cloud we calculate these timescales for all ions $i$ separately, as a local 
quantities within a cloud,
\begin{equation}
\label{eq:tion2}
 (t_{ion}^{i})^{-1} = 4 \pi \int_{\chi^i/h}^{+\infty} {J_{\nu} \over h \nu}\sigma^i(\nu)d\nu, 
\end{equation}
where $\sigma^i$ is the ionization cross-section and $\chi^i$ is the ionization energy of a given ion. Eq.~\ref{eq:tion} is derived analytically assuming a normal flux, neglecting the reflected flux (it is less than 10\% the incident flux), and using a classical formula for the cross-section (Osterbrock 1974).
The recombination timescale for each ion is estimated from the semi-analytical formulae 
of Verner \& Ferland (1996) 
\begin{equation}
 (t_{rec}^{i})^{-1} = n_e (\alpha_r(T) + \alpha_{diel}(T)),
\end{equation}
where $\alpha_r(T)$ is the radiative recombination coefficient and $\alpha_{diel}(T)$ is the coefficient of the dielectronic recombination.
Induced radiative recombination is negligible, hence this timescale does not depend on the flux.

As ionization or recombination timescales are specific of each ion, we analyzed the ionization structure of a whole cloud. An illustration of the method is shown in Table~\ref{tab:maxion_zones}, where we give the column densities, ionization and recombination timescales, zone per zone, for Fe{\sc xxv}--{\sc xxvii}, S{\sc xvi}--{\sc xvii}, Si{\sc xiv}--{\sc xv}, O{\sc viii}--{\sc ix} and C{\sc vi}--{\sc vii}, which are relevant for our model.
Recombination timescales for fully ionized elements are quite long (up to $2 \times 10^8$ s for hydrogen), but hydrogen and helium are almost completely ionized, and since these ions do not undergo frequent transitions, we determine the recombination timescale in each zone by identifying the dominant processes (i.e. both ions in interaction with this process must have roughly the same abundances). Although in Table~\ref{tab:maxion_zones} the longest recombination timescale is $6 \times 10^5$~s for the carbon in the hottest zone, but such transitions are of marginal importance. Therefore, the representative recombination timescale for a whole cloud is $2 \times 10^4$ s (again for Fe{\sc xxvi} in the hot zone). Representative ions for the recombination timescale in the intermediate, outer and end zones are Si{\sc xiv}, O{\sc ix} and C{\sc vii}, respectively.

Since these ionization and recombination timescales depend respectively on the incident flux and the density ($F = 8 \times 10^9$ erg s$^{-1}$cm$^{-2}$ and $n_o = 10^7$ cm$^{-3}$ in our reference cloud), we generalize the obtained values to Eqs.~\ref{eq:tion}, \ref{eq:trec}.
\begin{table}
\begin{tabular}{@{}cccccc@{}}
\hline
property & hot & intermediate & outer & end & all \\
\hline
\hline
Fe{\sc xxv} c & $5\ 10^{17}$ & $10^{18}$ & $2\ 10^{10}$ & 2 & $2\ 10^{18}$ \\
Fe{\sc xxvi} c & $7\ 10^{17}$ & $10^{17}$ & $4\ 10^7$ & $3\ 10^{-4}$ & $9\ 10^{17}$ \\
Fe{\sc xxvii} c & $6\ 10^{17}$ & $5\ 10^{15}$ & $3\ 10^4$ & $2\ 10^{-8}$ & $7\ 10^{17}$ \\
Fe{\sc xxvi} r & $2\ 10^4$ & 500 & 8 & 0.8 & $2\ 10^4$ \\
Fe{\sc xxvii} r & $10^4$ & 400 & 7 & 0.7 & $10^4$ \\
\hline
S{\sc xvi} c & $8\ 10^{16}$ & $2\ 10^{18}$ & $10^{15}$ & $2\ 10^{10}$ & $2\ 10^{18}$ \\
S{\sc xvii} c & $9\ 10^{17}$ & $10^{18}$ & $2\ 10^{13}$ & $2\ 10^7$ & $2\ 10^{18}$ \\
S{\sc xvi} i & 800 & $10^3$ & $10^3$ & $10^3$ & $10^3$ \\
\hline
Si{\sc xiv} c & $8\ 10^{16}$ & $3\ 10^{18}$ & $2\ 10^{16}$ & $2\ 10^{12}$ & $3\ 10^{18}$ \\
Si{\sc xv} c & $2\ 10^{18}$ & $4\ 10^{18}$ & $5\ 10^{14}$ & $5\ 10^9$ & $6\ 10^{18}$ \\
Si{\sc xv} r & $5\ 10^4$ & $2\ 10^3$ & 30 & 3 & $5\ 10^4$ \\
\hline
O{\sc viii} c & $3\ 10^{16}$ & $2\ 10^{18}$ & $3\ 10^{18}$ & $2\ 10^{17}$ & $7\ 10^{18}$ \\
O{\sc ix} c & $4\ 10^{19}$ & $10^{20}$ & $3\ 10^{18}$ & $7\ 10^{15}$ & $2\ 10^{20}$ \\
O{\sc viii} i  & 40 & 50 & 100 & 200 & 200 \\
O{\sc ix} r  & $2\ 10^5$ & $7\ 10^3$ & 100 & 10 & $2\ 10^5$ \\
\hline
C{\sc vi} c & $2\ 10^{15}$ & $10^{17}$ & $3\ 10^{17}$ & $6\ 10^{17}$ & $10^{18}$ \\
C{\sc vii} c & $2\ 10^{19}$ & $7\ 10^{19}$ & $3\ 10^{18}$ & $2\ 10^{17}$ & $10^{20}$ \\
C{\sc vi} i & 10 & 10 & 10 & 30 & 30 \\
C{\sc vii} r & $6\ 10^5$ & $10^4$ & 200 & 20 & $6\ 10^5$ \\
\hline
\end{tabular}
\caption{Table of the column densities (c, in cm$^{-2}$), ionization (i) and recombination (r) timescales (in s) for all 4 zones in the medium for the representative, i.e comparable column density for both ionization states, ions of Fe, S and Si, O, and C in the hot, intermediate, outer, and end zones, respectively.}
\label{tab:maxion_zones}
\end{table}


\begin{thebibliography}{99}

\bibitem[Allen(1973)]{1973asqu.book.....A} Allen, C.~W.\ 1973, London: 
University of London, Athlone Press, 1973, 3rd ed.

\bibitem[Ashton et al.(2006)]{2006MNRAS.366..521A} Ashton, C.~E., Page, 
M.~J., Branduardi-Raymont, G., \& Blustin, A.~J.\ 2006, \mnras, 366, 521

\bibitem[Bottorff et al.(2000)]{2000ApJ...537..134B} Bottorff, M.~C., 
Korista, K.~T., \& Shlosman, I.\ 2000, \apj, 537, 134

\bibitem[Blustin et al.(2005)]{2005A&A...431..111B} Blustin, A.~J., Page, 
M.~J., Fuerst, S.~V., Branduardi-Raymont, G., \& Ashton, C.~E.\ 2005, \aap, 
431, 111

\bibitem[Chelouche \& Netzer(2005)]{2005ApJ...625...95C} Chelouche, D., \& 
Netzer, H.\ 2005, \apj, 625, 95 

\bibitem[Collinge et al.(2001)]{2001ApJ...557....2C} Collinge, M.~J., et 
al.\ 2001, \apj, 557, 2 

\bibitem[Chevallier et al.(2006)]{2006A&A...449..493C} Chevallier, L., 
Collin, S., Dumont, A.-M., Czerny, B., Mouchet, M., Gon{\c c}alves, A.~C., 
\& Goosmann, R.\ 2006, \aap, 449, 493

\bibitem[Dumont et al.(2000)]{2000A&A...357..823D} Dumont, A.-M., 
        Abrassart, A., \& Collin, S.\ 2000, \aap, 357, 823 

\bibitem[Dumont et al. (2002)]{2002} Dumont, A.-M., Czerny, B., Collin, S.,
      \& \. Zycki, P.T., 2002, \aap, 387, 63

\bibitem[Dumont et al.(2003)]{2003A&A...407...13D} Dumont, A.-M., 
           Collin, S., Paletou, F., Coup{\' e}, S., Godet, O., \& Pelat, 
          D.\ 2003, \aap, 407, 13 

\bibitem[Elvis(2000)]{2000ApJ...545...63E} Elvis, M.\ 2000, \apj, 545, 63

\bibitem[Fabian et al.(2002)]{2002MNRAS.335L...1F} Fabian, A.~C., et al.\ 
2002, \mnras, 335, L1

\bibitem[Gabel et al.(2003)]{2003ApJ...595..120G} Gabel, J.~R., et al.\ 2003, \apj, 595, 120

\bibitem[Gierli{\'n}ski \& Done(2006)]{2006MNRAS.tmpL..64G} Gierli{\'n}ski, M., \& Done, C.\ 2006, \mnras, L64

\bibitem[Gon{\c c}alves et al.(2006)]{2006A&A...451L..23G} Gon{\c c}alves, 
A.~C., Collin, S., Dumont, A.-M., Mouchet, M., R{\'o}{\.z}a{\'n}ska, A., 
Chevallier, L., \& Goosmann, R.~W.\ 2006, \aap, 451, L23

\bibitem[Goosmann]{2006} Goosmann, R.W., Czerny, B., Mouchet, M. Ponti, G., Dovciak, M., Karas, V., Rozanska, A, Dumont, A.-M., 2006, A\&A in press (astro-ph/0604156)

\bibitem[Kaastra et al.(2002)]{2002A&A...386..427K} Kaastra, J.~S., 
Steenbrugge, K.~C., Raassen, A.~J.~J., van der Meer, R.~L.~J., Brinkman, 
A.~C., Liedahl, D.~A., Behar, E., \& de Rosa, A.\ 2002, \aap, 386, 427


\bibitem[Kaspi et al.(2002)]{2002ApJ...574..643K} Kaspi, S., et al.\ 2002, \apj, 574, 643 


\bibitem[Krolik et al.(1981)]{1981ApJ...249..422K} Krolik, J.~H., McKee, 
C.~F., \& Tarter, C.~B.\ 1981, \apj, 249, 422

\bibitem[Krolik \& Kriss(1995)]{1995ApJ...447..512K} Krolik, J.~H., \& Kriss, G.~A.\ 1995, \apj, 447, 512


\bibitem[Krongold et al.(2003)]{2003ApJ...597..832K} Krongold, Y., 
Nicastro, F., Brickhouse, N.~S., Elvis, M., Liedahl, D.~A., \& Mathur, S.\ 
2003, \apj, 597, 832 

\bibitem[Markowitz et al.(2003)]{2003ApJ...593...96M} Markowitz, A., et 
al.\ 2003, \apj, 593, 96

\bibitem[Murray \& Chiang(1995)]{1995ApJ...454L.105M} Murray, N., \& 
Chiang, J.\ 1995, \apjl, 454, L105

\bibitem[Murray \& Chiang(1997)]{1997ApJ...474...91M} Murray, N., \& 
Chiang, J.\ 1997, \apj, 474, 91 

\bibitem[Murray \& Chiang(1998)]{1998ApJ...494..125M} Murray, N., \& 
Chiang, J.\ 1998, \apj, 494, 125

\bibitem[Netzer(1993)]{1993ApJ...411..594N} Netzer, H.\ 1993, \apj, 411, 594

\bibitem[Netzer(1996)]{1996ApJ...473..781N} Netzer, H.\ 1996, \apj, 473, 
781

\bibitem[Netzer et al.(2003)]{2003ApJ...599..933N} Netzer, H., et al.\ 
2003, \apj, 599, 933

\bibitem[Ogle et al.(2004)]{2004ApJ...606..151O} Ogle, P.~M., Mason, K.~O., 
Page, M.~J., Salvi, N.~J., Cordova, F.~A., McHardy, I.~M., \& Priedhorsky, 
W.~C.\ 2004, \apj, 606, 151 

\bibitem[Osterbrock (1974)]{1974} Osterbrock, D.E., 1974, Astrophysics of Gaseous Nebulae, 
W.H.Freeman and Company, San Francisco


\bibitem[Papadakis(2004)]{2004MNRAS.348..207P} Papadakis, I.~E.\ 2004, \mnras, 348, 207


\bibitem[Piconcelli et al.(2004)]{2004MNRAS.351..161P} Piconcelli, E., 
Jimenez-Bail{\'o}n, E., Guainazzi, M., Schartel, N., 
Rodr{\'{\i}}guez-Pascual, P.~M., \& Santos-Lle{\'o}, M.\ 2004, \mnras, 351, 
161 

\bibitem[Pietrini \& Torricelli-Ciamponi(2000)]{2000A&A...363..455P} 
Pietrini, P., \& Torricelli-Ciamponi, G.\ 2000, \aap, 363, 455 

\bibitem[Ponti et al.(2004)]{2004A&A...417..451P} Ponti, G., Cappi, M., Dadina, M., \& Malaguti, G.\ 2004, \aap, 417, 451

\bibitem[Proga et al.(2000)]{2000ApJ...543..686P} Proga, D., Stone, J.~M., 
\& Kallman, T.~R.\ 2000, \apj, 543, 686 

\bibitem[Proga \& Kallman(2004)]{2004ApJ...616..688P} Proga, D., \& 
Kallman, T.~R.\ 2004, \apj, 616, 688 

\bibitem[R{\'o}{\.z}a{\'n}ska et al.(2006)]{2006A&A...452....1R} 
R{\'o}{\.z}a{\'n}ska, A., Goosmann, R., Dumont, A.-M., \& Czerny, B.\ 2006, 
\aap, 452, 1 

\bibitem[Rybicki \& Hummer(1991)]{1991A&A...245..171R} Rybicki, G.~B., \& 
Hummer, D.~G.\ 1991, \aap, 245, 171 

\bibitem[Sako et al.(2002)]{2002xsac.conf..191S} Sako, M., et al.\ 2002, 
X-ray Spectroscopy of AGN with Chandra and XMM-Newton, 191 

\bibitem[Scott et al.(2005)]{2005ApJ...634..193S} Scott, J.~E., et al.\ 
2005, \apj, 634, 193 

\bibitem[Smith \& Raine(1988)]{1988MNRAS.234..297S} Smith, M.~D., \& Raine, 
D.~J.\ 1988, \mnras, 234, 297 

\bibitem[Steenbrugge et al.(2005)]{2005A&A...434..569S} Steenbrugge, K.~C., et al.\ 2005, \aap, 434, 569

\bibitem[Torricelli-Ciamponi \& Courvoisier(1998)]{1998A&A...335..881T} 
Torricelli-Ciamponi, G., \& Courvoisier, T.~J.-L.\ 1998, \aap, 335, 881

\bibitem[Verner \& Ferland(1996)]{1996ApJS..103..467V} Verner, D.~A., \& Ferland, G.~J.\ 1996, \apjs, 103, 467 


\end{thebibliography}
\end{document}